\documentclass{aa} 
\pdfoutput=1
\usepackage{graphicx}
\usepackage{txfonts}
\usepackage{url}
\usepackage{amsmath}
\usepackage{natbib,twoopt}
\usepackage[breaklinks=true]{hyperref} 
\hypersetup{colorlinks=true,urlcolor=blue,citecolor=blue,pdfborder= 0 0 0}
\bibpunct{(}{)}{;}{a}{}{,}    
\makeatletter
 \newcommandtwoopt{\citeads}[3][][]{\href{http://adsabs.harvard.edu/abs/#3}%
 {\def\hyper@linkstart##1##2{}%
  \let\hyper@linkend\@empty\citealp[#1][#2]{#3}}}
 \newcommandtwoopt{\citepads}[3][][]{\href{http://adsabs.harvard.edu/abs/#3}%
 {\def\hyper@linkstart##1##2{}%
  \let\hyper@linkend\@empty\citep[#1][#2]{#3}}}
 \newcommandtwoopt{\citetads}[3][][]{\href{http://adsabs.harvard.edu/abs/#3}%
 {\def\hyper@linkstart##1##2{}%
  \let\hyper@linkend\@empty\citet[#1][#2]{#3}}}
 \newcommandtwoopt{\citeyearads}[3][][]%
 {\href{http://adsabs.harvard.edu/abs/#3}
 {\def\hyper@linkstart##1##2{}%
  \let\hyper@linkend\@empty\citeyear[#1][#2]{#3}}}
\makeatother

\usepackage[breaklinks=true]{hyperref} 
\hypersetup{colorlinks=true,urlcolor=blue,citecolor=blue,pdfborder= 0 0 0}
\begin{document}

 \title{From voids to filaments: \\ environmental transformations of galaxies in the SDSS}

 \author{Teet Kuutma\inst{1,2}
          \and Antti Tamm\inst{1}
          \and Elmo Tempel\inst{1,3}
   }

 \institute{Tartu Observatory, Observatooriumi~1, 61602 T\~oravere, Estonia\\
   \email{teet.kuutma@to.ee}
   \and
   Institute of Physics, University of Tartu, Ostwaldi~1, 50411 Tartu, Estonia
   \and
   Leibniz-Institut für Astrophysik Potsdam (AIP), An der Sternwarte~16, 14482 Potsdam, Germany
 }

 \date{}

 
 \abstract
 {} 
 {We investigate the impact of filament and void environments on galaxies, looking for residual effects beyond the known relations with environment density.}
 {We quantified the host environment of galaxies as the distance to the spine of the nearest filament, and compared various galaxy properties within 12 bins of this distance. We considered galaxies up to 10 $h^{-1}$Mpc from filaments, i.e. deep inside voids. The filaments were defined by a point process (the Bisous model) from the Sloan Digital Sky Survey data release 10. In order to remove the dependence of galaxy properties on the environment density and redshift, we applied weighting to normalise the corresponding distributions of galaxy populations in each bin.}
 {After the normalisation with respect to environment density and redshift, several residual dependencies of galaxy properties still remain. Most notable is the trend of morphology transformations, resulting in a higher  elliptical-to-spiral ratio while moving from voids towards filament spines, bringing along a corresponding increase in the $g-i$ colour index and a decrease in star formation rate. After separating elliptical and spiral subsamples, some of the colour index and star formation rate evolution still remains. The mentioned trends are characteristic only for galaxies brighter than about $M_{r} = -20$ mag. Unlike some other recent studies, we do not witness an increase in the galaxy stellar mass while approaching filaments. The detected transformations can be explained by an increase in the galaxy-galaxy merger rate and/or the cut-off of extragalactic gas supplies (starvation) near and inside filaments.}
 {Unlike voids, large-scale galaxy filaments are not a mere density enhancement, but have their own specific impact on the constituent galaxies, reducing the star formation rate and raising the chances  of elliptical morphology also at a fixed environment density level.}


 \maketitle
%
\section{Introduction}

Large-scale structure filaments, the elongated chains of galaxies and galaxy groups framing the huge underdense voids, are among the most eye-catching features of the cosmic web. They are products of the anisotropic character of the gravitational force field in a random matter distribution. The resulting tidal forces typically lead to anisotropic contraction and collapse of matter into wall-like and elongated filamentary structures \citep[][]{Zeldovich:70, Bond:96, vandeWeygaert:08}. The filament-void pattern is well reproduced in cosmological simulations \citep[][]{Doroshkevich:80, Klypin:83, Davis:85, Gramann:93, Sheth:04, Springel:05}.
In recent years, filaments have been found to occur over the wide span of large-scale density levels, residing inside voids and  superclusters \citep{Aragon-Calvo:10, Sousbie:11, Einasto:12, Alpaslan:14, Cautun:14, Darvish:16, Malavasi:17}.

Observations of prominent filaments \citep[][]{Giovanelli:86,Ebeling:04, Porter:05} and detailed numerical simulations \citep[][]{Colberg:05, Benitez-Llambay:13, Cautun:14, Tempel:14c} indicate that filaments are not only a manifestation of matter distribution, but also a marker of a particular dynamical phase of the large-scale structure network in which matter is channelling towards massive clusters. Therefore, we can expect that as a distinct environment with its own dynamics and matter flows, filaments should have some effect on their constituent galaxies. 

The main differences between void galaxies and filament galaxies, the former being generally less massive, bluer, and more gas rich \citep{Kreckel:11,Rojas:04,Hoyle:05,Beygu:17}, are primarily characterised by the well-established morphology--density relation \citep{Hubble:31, Dressler:80}. Additionally, tidal torques are expected to add angular momentum to galaxies and galaxy systems inside filaments, due to the highly inhomogeneous large-scale matter distribution \citep[e.g.][]{Lee:00}.

From observations, various hints for a generally coherent intrinsic dynamics of filaments can be found from several recent alignment studies. Namely, galaxies \citep{Jones:10, Tempel:13a, Tempel:13b, Hirv:16, Pahwa:16, Rong:16}, galaxy pairs \citep{Tempel:15b}, and larger satellite systems \citep{Tempel:15a} have been reported to align with cosmic filaments to a notable extent. This can be explained by the velocity field being dominantly parallel to filaments \citep{Tempel:14c, Libeskind:15}, coupled with the momentum conservation principle.

Apart from the alignment indicators, there is some observational evidence of additional filament-induced changes in galaxy populations. Stellar mass and star formation rate as a function of galaxy distance from filaments has been studied using the GAMA survey data \citep{Alpaslan:15, Alpaslan:16}. In isolated spiral galaxies it was found that  stellar mass increases, but star formation per unit stellar mass (the specific star formation rate, sSFR) decreases closer to filaments. The same effect has been detected at $z\approx0.7$ in the VIPERS survey \citep{Malavasi:17}. From a theoretical point of view, \citet{Aragon-Calvo:16} showed that the star formation quenching in galaxies can be explained as the influence of filamentary environment, due to the non-linear interaction of the cosmic web. Meanwhile, a variety of case studies of  particular filamentary structures have reported an increased fraction of star-forming galaxies \citep{Fadda:08, Tran:09, Biviano:11, Darvish:14}, and higher metallicities and lower electron densities \citep{Darvish:15} in filaments with respect to field environments. Unfortunately, most of these studies encompass a relatively small number of galaxies and limited volumes, making them sensitive to the cosmic variance.

So far, the lush dataset of the SDSS offers an optimal compromise between the number density of objects and the covered volume for comparing void and filament environments. However, it has not yet been exploited in its full power for this purpose. \citet{Poudel:17} find that in filaments, galaxy groups tend to be richer and their central galaxies preferentially of earlier types than in voids, while \citet{Shim:15} have detected an interesting relation between galaxy luminosity and the straightness of their host filaments.

In this paper we make use of the SDSS dataset to probe galaxy properties as a function of distance to the nearest filament spine, relying on the large collection of filaments catalogued by \citet{Tempel:14a}. We are interested in distinguishing low-level phenomena (the coherent velocity field inside filaments, details of gasodynamics, etc.) from the well-known impact of environment density, thus we normalise our galaxy samples with respect to the 1 $h^{-1}$Mpc smoothed environment density level.

Throughout this paper we assume the following cosmology: the Hubble constant $H_0 = 70~\mathrm{km~s^{-1}Mpc^{-1}}$, the matter density $\Omega_\mathrm{m} = 0.27$, and the dark energy density $\Omega_\Lambda = 0.73$.

\section{Data and methods}

\subsection{Galaxy sample}
The subsequent analysis is based on the SDSS data release~10 \citep{York:00, Ahn:13}. The galaxy sample was refined as described in \citet{Tempel:14d}, removing spurious detections and galaxies with unreliable parameters. Instead of using flux-limited samples, where the number density of galaxies decreases with distance, smaller volume-limited samples were used in the core part of the study, ensuring a uniform completeness level over the probed distance range. We considered two subsamples: a brighter one ($M_r < -20$ mag) and a fainter one ($-18$ mag $> M_r > -20$ mag). The samples cover distance ranges of $11 < h^{-1}$Mpc $< 320$ and $6 < h^{-1}$Mpc $< 135$, respectively.

In order to measure and characterise the impact of filaments, we considered the elliptical-to-spiral (E/S) ratio and the mean broad-band ${g-i}$ colour index, star formation rate, concentration index, and the stellar velocity dispersions of galaxy populations as a function of distance to the nearest filament spine. To probe galaxy morphology, we used visual estimates from the Galaxy Zoo program \citep{Lintott:08} and algorithm-based morphology estimates from \citet[][hereafter HC11]{Huertas-Company:11}.

We used the galaxy data  available on the SDSS SkyServer. The $g$ and $i$ luminosities and stellar velocity dispersions are those from the SDSS team, the concentration index was calculated as the ratio of the Petrosian radii $R_{50}/R_{90}$. The stellar mass and star formation estimates are those of the Granada group \citep{Conroy:09};  of their four different models, the one yielding the lowest $\chi^2$  fitting was selected for each galaxy.

\subsection{Filament sample}
To quantify the large-scale environment of a galaxy, we used its distance to the nearest filament spine (designated as $D_{\rm{fil}}$ hereafter). We used filaments as extracted from the SDSS data via an object point process, described in detail in \citet{Tempel:14a} and \citet{Tempel:16a}. In brief, filamentary structures were probed with cylindrical segments, which were fitted to and adjusted on the three-dimensional galaxy distribution. The process is stochastic, so variations between different Markov chain Monte Carlo runs of the fitting reveal the likelihood of the detected filaments. The axes of sequential cylinders form the spine of a filament. The filaments in the given catalogue were probed using cylinders of 1 $h^{-1}$Mpc diameter;   at this scale, filaments appear to be the most robust, at least in the case of the SDSS dataset. We repeated the study presented below  using 2 $h^{-1}$Mpc and 4 $h^{-1}$Mpc radius filaments, revealing quantitatively similar, but mildly weaker trends.

In general, since the distances of galaxies are redshift-based, any three-dimensional large-scale structure mapping is affected by the redshift-space distortions, especially the Fingers-of-God effect arising from the peculiar velocities of galaxies, and the Kaiser effect \citep{Kaiser:87} caused by large-scale matter flows. In this work, instead of using the redshift of each individual galaxy to derive its distance, we used the distance of the centre of the group a galaxy belongs to, calculated as the average over all galaxies within the group. Group membership (determined with the friends-of-friends algorithm) and the corresponding distance are taken from the galaxy group catalogue \citet{Tempel:14d}. With this distance definition we are able to suppress the Fingers-of-God effect, enabling us to delineate filaments and measure galaxy distances from it more reliably. We also tested  for the Kaiser effect by comparing the results for filaments with axes along the line of sight to those along the sky plane. No significant differences were found; thus, we can assume that the Kaiser effect does not play a significant role and as such is safely  ignored in the scope of this study. At intermediate scales, the potential impact of infall motions towards clusters, which may cause additional redshift-space distortions, is  reduced first by the filament finder, which is specifically tuned to avoid the inclusion of clusters, and then by our use of filaments longer that 15 $h^{-1}$Mpc, thus further restricting ourselves to only well-defined filaments.

We considered galaxies located up to 10 $h^{-1}$Mpc from the nearest filament spine. In the local universe, the uncertainties of galaxy distances and the large angular size makes the detected filaments uncertain; thus, below 60 $h^{-1}$Mpc (corresponding to $z \approx 0.02$), a distance cut-off was applied.

The filament detection method is based on comparing the galaxy number density inside and outside the cylindrical elements. The number of galaxies in the immediate vicinity of the cylinder boundaries is low, yielding large statistical errors. Therefore, we have considered separately galaxy populations inside the filaments (0 -- 0.5 $h^{-1}$Mpc from the spine) and outside the filaments (0.7 -- 10 $h^{-1}$Mpc from the spine). Given that we also divided the galaxy sample into a brighter and a fainter part, we ended up with four subsamples. For each of the subsamples, mean galaxy properties were calculated within 6 bins of distance from the filament spine. The bin boundaries were chosen so that all bins would contain roughly equal numbers of galaxies and their placement would evenly cover the logarithmic distance scale.

\subsection{Normalising the galaxy populations}\label{sect:norm}

The general environment density increases while moving from voids towards the spines of filaments, thus the constituent galaxy populations are correspondingly affected by the established relations between galaxy properties and environment density \citep{Dressler:80,Kauffmann:04}, which have been shown to act over a wide range of environments and spatial scales \citep{Park:07, Lietzen:12, Einasto:14,Poudel:16a}. However, our aim here is to look for differences between void and filament environments beyond the known dependencies on environment density.

Another aspect to be taken into account is  that several measured properties of galaxies are systematically dependent on distance (redshift), due to observational limitations. Since filament properties are also a function of distance because of the changing galaxy population (the filaments are based on flux-limited data to increase the number of detected structures), there is a danger of introducing a distance-related bias into the results.

\begin{figure}
 \centering
 \includegraphics[width=88mm]{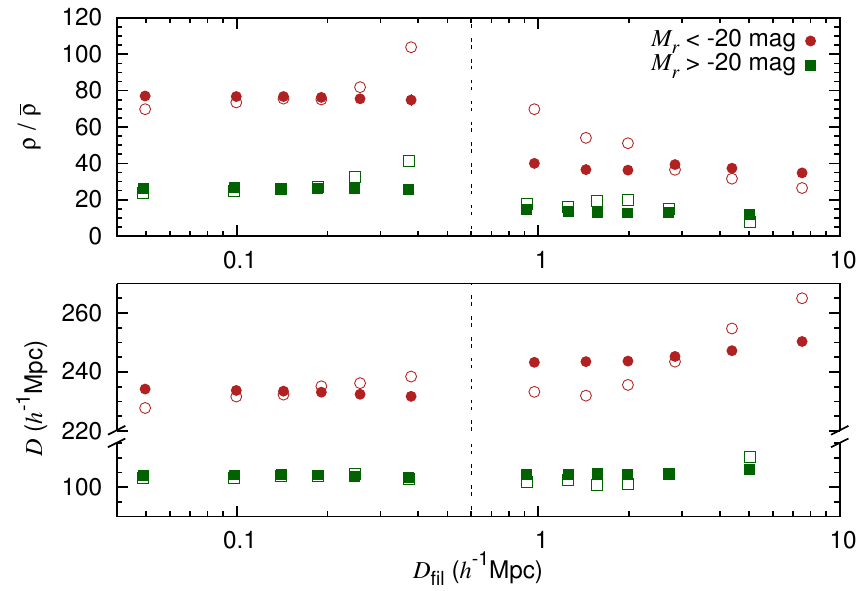}
 \caption{Mean environment density around the galaxies (upper panel) and the mean distance of galaxies (lower panel) as a function of distance from the nearest filament spine, $D_{\rm{fil}}$. The empty (filled) symbols correspond to the values before (after) normalisation. Red circles represent the brighter volume-limited sample and green squares the fainter one. The 2$\sigma$ variations derived from 1000 bootstrap sampling iterations are smaller than the points on the graph.}
 \label{fig:dendist}
\end{figure}

The mean environment density and distance of galaxies inside each $D_{\rm{fil}}$ bin are shown in Fig.~\ref{fig:dendist}. Here, the empty red circles indicate the mean value for each bin for the brighter sample and the empty green squares for the fainter sample. As the environment density ($\rho/\overline{\rho}$, in units of the mean of the Universe), we utilised the $r$-band luminosity density smoothed over 1 $h^{-1}$Mpc scale with a $B_{3}$ spline kernel, given in \citet{Tempel:14d}, after subtracting the contribution by a given galaxy. We see that galaxies closer to filaments tend to be found in denser environments and at lower redshifts than the ones farther in the voids. The turnaround of the density trend inside the filaments is a result of the detection of weaker filaments in the nearby Universe.

To neutralise the  effects imposed by environment density and distance, we have normalised the galaxy populations inside each $D_{\rm{fil}}$ bin with respect to density and distance by applying a corresponding weight to each galaxy. As the first step of normalisation, we made sure  that all $D_{\rm{fil}}$ bins covered the same range of environment density and distance by excluding any galaxy lying outside the Gaussian-smoothed common range of densities and distances. This criterion removed $0-7\%$ of the galaxies, depending on the subsample. The remaining galaxies inhabit environments with densities $\rho/\overline{\rho}\leq900$ and $\rho/\overline{\rho}\leq300$ for the brighter and fainter sample, respectively. At low densities, $\rho/\overline{\rho}$ becomes uncertain, thus we applied a  cut-off at $\rho/\overline{\rho} = 0.1$ for both samples.

The weights of the galaxies were calculated separately for each of the four galaxy subsamples, taking the total distribution of the subsample as the reference. For more than 97\% of the galaxies, the weights remained between 0.5 and 2. After the refinement of the subsamples, each $D_{\rm{fil}}$ bin contained about 5000 galaxies in the brighter sample and about 900 galaxies in the fainter sample. The results of the normalisation are shown in Fig.~\ref{fig:dendist} with filled symbols. It is evident that the weighting has removed all dependence on environment density and distance. Thus, any trends in the subsequent analysis are detached from these effects. We note that the distributions for the inner and outer subsamples (inside and outside the 0.6 $h^{-1}$Mpc radius, indicated by the vertical dashed line) should not be compared directly to each other because of a slightly different normalisation.

We also tested the effect of constraining the environment densities by using only isolated galaxies on the basis of \citet{Tempel:14d}. The final results were almost identical, but the uncertainties were larger, due to a smaller number of galaxies. Thus we can conclude that a careful normalisation of the galaxy sample according to environment density is sufficient and yields better statistics than the isolation criterion.

\begin{figure}
 \centering
 \includegraphics[width=88mm]{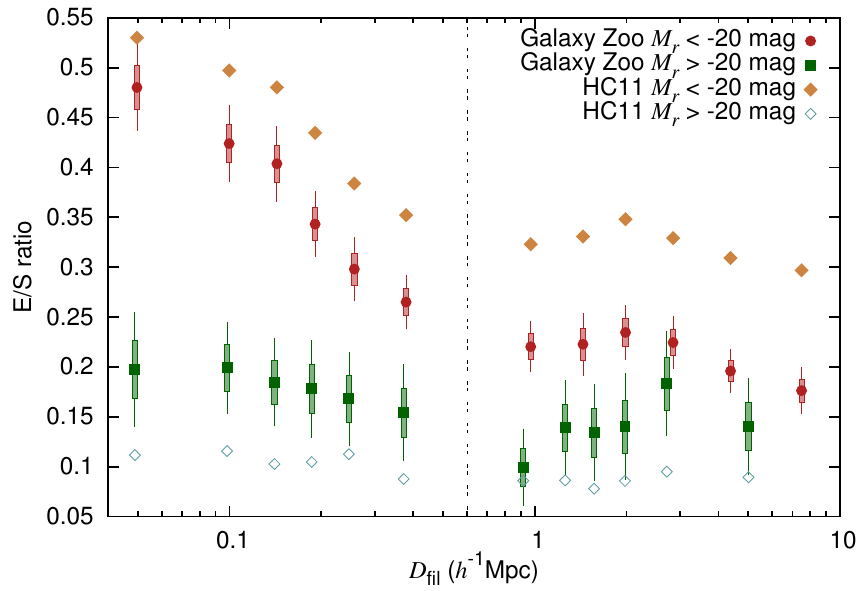}
 \caption{Average E/S ratios as a function of the distance $D_{\rm{fil}}$. Galaxy Zoo morphologies are represented by red circles for the brighter sample and by green squares for the fainter sample. Classifications from HC11 are shown by filled yellow diamonds for the brighter sample and by empty blue diamonds for the fainter sample. The bootstrap uncertainties for the HC11 dataset are smaller than the points on the graph.}
 \label{fig:es}
\end{figure}

Galaxy properties are also known to be strong functions of the stellar mass \citep[][]{Kauffmann:03}. \citet{Alpaslan:16} showed that galaxy stellar mass increases while approaching the filament axis. In our case, with the use of volume-limited samples and correspondingly narrower mass ranges, this effect is significantly reduced. Moreover, we found that the average stellar mass changes very little with the distance $D_{\rm{fil}}$ even when using a large flux-limited galaxy sample, and any additional weighting by stellar mass has no effect on the outcome.

\section{Results and discussion}\label{sect:res}

After the normalisations described in the previous section, we calculated the mean properties of the galaxy populations for each $D_{\rm{fil}}$ bin. In Fig.~\ref{fig:es}, the distribution of the ratios of galaxy morphologies as classified by the Galaxy Zoo project and HC11 are presented; the 1$\sigma$ and 2$\sigma$ bootstrap errors are also shown. We see that early-type galaxies are more abundant closer to filament axes than farther away. Quantitatively, this is similar to the morphological segregation in the Pisces-Perseus supercluster shown already by \citet{Giovanelli:86}. However, in Pisces-Perseus, the effect can be explained by the known morphology-density relation, while in our case, this relation has been neutralised and we are witnessing a trend on top of that. The trend is especially strong inside the filaments. For the low-luminosity galaxies, no clear trend is visible beyond the uncertainties; interestingly, however, there is a vague hint of a slight increase in the spiral fraction just off the edge of the filament for both samples, which is in agreement with the increase in \ion{H}{I} supplies near filaments suggested by \citet{Kleiner:16}.

\begin{figure}
 \centering
 \includegraphics[width=88mm]{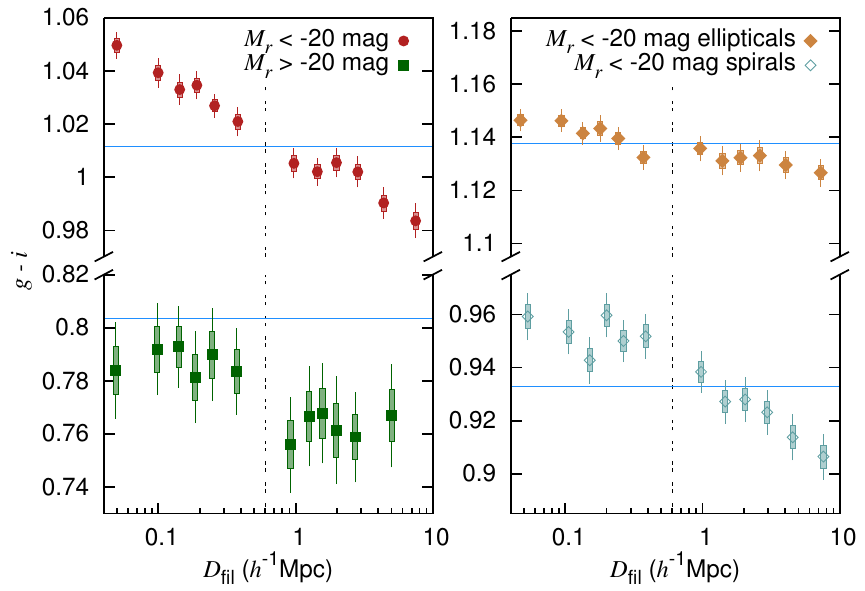}
 \caption{Average $g-i$ values  as a function of the distance $D_{\rm{fil}}$. In the left panel red circles are for the brighter sample  and green squares for the fainter sample. In the right panel there are separate  distributions  for the brighter sample spiral and elliptical galaxies according to the HC11 classification. Yellow diamonds indicate ellipticals and blue diamonds indicate spirals. The horizontal blue lines represent the average $g-i$ values for the corresponding volume-limited samples.}
 \label{fig:gi}
\end{figure}

The morphological transformation brings  an expected trend of the mean colour index ${g-i}$, plotted in the left panel of Fig.~\ref{fig:gi}, and a corresponding change in the sSFR (not plotted). Closer to filaments and closer to filament spines within filaments, brighter galaxies tend to become redder, while fainter galaxies display no clear trend beyond the error bars. For comparison, the horizontal lines indicate the average $g-i$ value for the whole volume-limited sample (galaxies in all environments) within the corresponding luminosity range. Similar  trends occur in the concentration index and the stellar velocity dispersion; being easily predictable from the morphology evolution, we omit the plots here.

While it is clear that the morphology transformation is responsible for the bulk of the change in the colour index and in the sSFR, it is also of fundamental importance whether any change remains also after separating the elliptical and spiral subpopulations. Splitting the brighter sample according to the HC11 classification (and skipping the fainter one because of the large uncertainties), we see that both groups still experience reddening with decreasing $D_{\rm{fil}}$ with a stronger trend for spiral galaxies. After splitting according to morphology there is almost no trend in the concentration index. However, some development takes place in the stellar velocity dispersion, where within filaments, elliptical galaxies (and to a lesser extent spiral galaxies)  closer to the spine become dynamically hotter.

The detected increase in the colour index and the decrease in the sSFR with $D_{\rm{fil}}$ are qualitatively similar to the results derived for the spiral galaxies in the GAMA sample by \citet{Alpaslan:16}. However, it is important to note that in the latter work,  a mass evolution trend was also reported, and more massive galaxies tend to have reduced sSFR. In contrast, we find that sSFR remains lower at a fixed stellar mass; a similar result has been reached for the bright galaxies in the VIPERS survey \citep{Malavasi:17}.

Taken together, we can conclude that filaments do not differ from voids only in their general density, but also due to some endemic processes. Even at a fixed stellar mass, the likelihood that a massive galaxy is a star-forming spiral decreases while approaching filaments and filament spines within filaments,  a trend probably related to an enhanced merger rate and/or to the cut-off of external gas supplies \citep{Aragon-Calvo:16}. Inside filaments, massive elliptical galaxies become dynamically hotter closer to filament spines, indicating a further increase in the merger rate and/or satellite accretion near the spines. Since the environmental density level has been kept fixed in our study, the possible increase in the merger rate can only be due to higher peculiar velocities of the galaxies or due to the increase in satellite number density close to filament spines \citep{Guo:15}.

In the case of the lower-mass galaxy sample, no evolution from voids to filament spines is witnessed. However, since the sample is considerably smaller than the higher mass one, potential weaker trends may be overwhelmed by uncertainties. On the other hand, an apparent non-evolution of this population may be a result of  a missing evolution, possibly explained by the fact that these galaxies inhabit somewhat lower density environments, or a result of  an evolution together with a luminosity increase, causing a shift of the galaxies into the brighter sample.

\begin{acknowledgements}

  We thank the referee for the constructive comments which have helped to improve this paper.
  
  This work was supported by grants IUT26-2 and IUT40-2 of the Estonian Ministry of Education and Research and the Centre of Excellence ``Dark side of the Universe'' (TK133), financed by the European Regional Development Fund. 
  
  Funding for SDSS-III has been provided by the Alfred P. Sloan Foundation, the Participating Institutions, the National Science Foundation, and the U.S. Department of Energy Office of Science. The SDSS-III web site is http://www.sdss3.org/. 

\end{acknowledgements}

\bibliographystyle{aa} 
\bibliography{mybib} 

\end{document}